\documentclass[twocolumn,twoside,slac_two]{revtex4}
\usepackage{graphicx}
\usepackage{fancyhdr}
\begin{document}

\title{The Search for Neutrinoless Double Beta Decay in CUORE}

%

\author{L. M. Ejzak \small{on behalf of the CUORE collaboration}}
\affiliation{Department of Physics, University of Wisconsin-Madison, Madison, WI 53706, USA}

\begin{abstract}
Understanding the nature of neutrino masses will require physics beyond the long-standing Standard Model of particle physics. Neutrinoless double beta decay ($0\nu\beta\beta$) experiments like the Cryogenic Underground Observatory for Rare Events (CUORE) are uniquely suited for probing the remaining mysteries of neutrino mass, particularly the question of the neutrino's Majorana nature. CUORE will be a next-generation experiment at the Laboratori Nazionali del Gran Sasso in Italy; it will consist of an array of 988 TeO$_2$ detector crystals operated at~$\sim$10~mK, following the bolometric technique established by the Cuoricino experiment. It will look for the energy signal produced by the theoretically-predicted $0\nu\beta\beta$ decay in $^{130}\mbox{Te}$, and therefore reliable energy calibration of the detector is crucial to the experiment's success. We will present the most recent results from Cuoricino and discuss the current status of the CUORE project, with a particular emphasis on the development of the calibration system.
\end{abstract}

\maketitle

\thispagestyle{fancy}


\section{Introduction}

Nuclear beta decay is the familiar process in which a nucleon decays, releasing a positron (electron) and an electron neutrino (antineutrino). For some even-even nuclei of mass A and charge Z, beta decay is forbidden, since the resulting (A,Z$\pm$1) nucleus would be less bound than the initial nucleus; however, if there is a (A,Z$\pm$2) nucleus which is more bound than the initial nucleus, it is possible to observe the process called two-neutrino double beta decay ($2\nu\beta\beta$), in which two nucleons decay simultaneously. The half-lives for $2\nu\beta\beta$ are long - typically of order $10^{18}-10^{21}$ years, and some are even longer - since it is a second-order weak process, but this decay has been observed in a number of isotopes \cite{Barabash}. The sum energy spectrum of the emitted betas is a continuous distribution between zero and the Q-value of the decay, since the neutrinos carry off some of the energy.

If neutrinos are Majorana particles, there is the possibility that, some small fraction of the time, these isotopes could undergo neutrinoless double beta decay ($0\nu\beta\beta$) instead, in which the two (anti)neutrinos would disappear in a virtual particle exchange instead of being released as physical particles. The signal for this process is a sharp peak in the beta energy spectrum right at the Q-value of the decay, as the nucleus is so heavy that the recoil is negligible. At this time, the only feasible experimental approach to determining whether or not neutrinos are Majorana particles is to search for evidence of $0\nu\beta\beta$. There are a number of questions about neutrino properties that $0\nu\beta\beta$ has the potential to address:
\begin{enumerate}
 \item Whether neutrinos are Majorana particles
 \item Absolute neutrino mass scale
 \item Neutrino mass hierarchy
\end{enumerate}

The reason that $0\nu\beta\beta$ may be able to provide a handle on one or more of these questions is that, in the case that $0\nu\beta\beta$~is mediated by the exchange of a virtual light Majorana neutrino, the $0\nu\beta\beta$ decay rate $\Gamma_{0\nu}$ is related to the neutrino mass. To be more specific,
\begin{equation}
 \Gamma_{0\nu}=G^{0\nu}\arrowvert M_{nucl}\arrowvert^{2}\langle m_{\nu}\rangle^{2},
 \label{eq:rate}
\end{equation}
\noindent
where the effective $0\nu\beta\beta$ neutrino mass is
\begin{equation}
 \langle m_{\nu}\rangle = \left|\sum_{j}\arrowvert U_{ej}\arrowvert^{2}e^{i\phi_{j}}m_{j}\right|.
 \label{eq:m_nu}
\end{equation}
In the above equations, $G^{0\nu}$ is a phase space integral, $M_{nucl}$ represents nuclear matrix elements, the $U_{ej}$ are elements of the neutrino mixing matrix, the $\phi_{j}$ are possible complex Majorana phases, and the $m_{j}$ are the physical neutrino mass eigenvalues \cite{Elliot}. Experimental results are often expressed in terms of the half life $T^{0\nu}_{1/2}$~($\propto\Gamma_{0\nu}^{-1}$).

Unfortunately, a number of complications arise here that limit how much the quantity that is actually being measured, $\Gamma_{0\nu}$, can say about the quantities of interest. While $G^{0\nu}$ can be accurately calculated, various theoretical calculations of the nuclear matrix elements can differ by a factor of 2-3, meaning there is significant theoretical uncertainty in determining the effective mass from the decay rate, and additional difficulty in using the effective mass to set constraints on the physical masses comes from the fact that the phases $\phi_{j}$ are completely unknown \cite{McKeown}.

Nonetheless, it is possible to use the information about the neutrino mixing parameters that has been gained from oscillation experiments to determine the allowed phase space for $\langle m_{\nu}\rangle$ with respect to the mass of the lightest neutrino mass state. The allowed region splits into two distinct regions corresponding to the normal hierarchy and the inverted hierarchy. This is what gives $0\nu\beta\beta$ experiments some discriminatory power when it comes to the mass hierarchy.

It is important to note that the above discussion applies \emph{only} in the simple case that $0\nu\beta\beta$ is mediated by the exchange of a virtual light Majorana neutrino (which will be assumed for the rest of this paper). If some other lepton-number-violating physics is responsible (e.g., exchange of a heavy Majorana neutrino or supersymmetric particles), it becomes much more difficult if not impossible to extract $\langle m_{\nu}\rangle$ \cite{Elliot}. However, no matter what mechanism is responsible, if $0\nu\beta\beta$ is observed then neutrinos are Majorana particles \cite{Schechter}.

\section{The Experimental Situation}

Since $0\nu\beta\beta$, assuming it happens at all, is such a rare process - even more so than $2\nu\beta\beta$, which, as discussed previously, has quite a long half life to begin with - the attempt to observe it experimentally poses a challenge. One way to characterize the sensitivity of an experiment is to introduce a figure of merit F, defined as the ratio of the number of signal events to the Poisson fluctuation of the background:
\begin{equation}
 F=\frac{N_{s}}{\sqrt{N_{b}}}=\Gamma_{0\nu}\cdot f \cdot\epsilon\cdot\sqrt{\frac{N\cdot\mathcal{T}}{\Delta E\cdot b}}
\label{eq:sensitivity}
\end{equation}
\noindent
where $f$ is the isotopic fraction, $\epsilon$ is the detector efficiency, $N$ is the total number of nuclei (i.e., detector mass), $\mathcal{T}$ is the live time of the experiment, $\Delta E$ is the energy resolution of the detector, and $b$ is the constant background rate per atom per energy interval~\cite{Piquemal}. Thus a successful $0\nu\beta\beta$ experiment will need to be a long-running, low-background experiment with good resolution and a large-mass detector.

\begin{figure}[b]
 \centering
 \includegraphics[width=80mm]{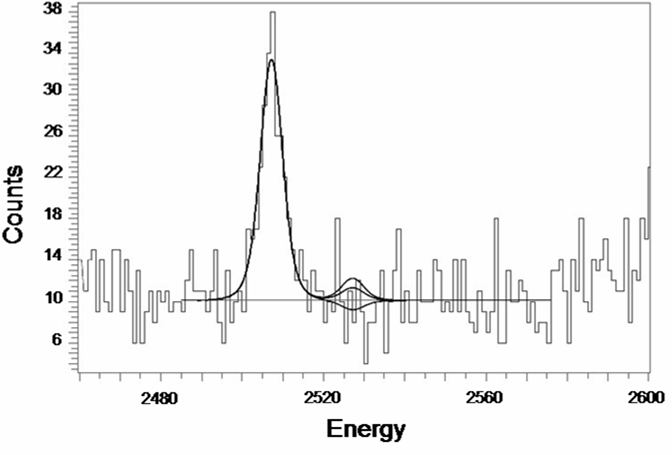}
 \caption{The most recent evaluation of the Cuoricino limit (results are preliminary), representing an exposure of $18\mbox{ yr}\cdot\mbox{kg }^{130}\mbox{Te}$. The left-hand peak comes from gammas caused by $^{60}\mbox{Co}$ contamination resulting from cosmogenic activation of the detector's copper support structures; it is close enough to the expected Q-value that it must be included in the $0\nu\beta\beta$ fit. The three $0\nu\beta\beta$ contours shown are, from bottom to top: best fit, 1$\sigma$, and 90\% C.L.}
 \label{fig:limit}
\end{figure}

At the moment, experiments using $^{76}\mbox{Ge}$ (Heidelberg-Moscow, IGEX) and $^{130}\mbox{Te}$ (Cuoricino) hold the best limits on $\langle m_{\nu}\rangle$~\cite{Piquemal}. These limits lie in the so-called degenerate-hierarchy region, where the absolute neutrino mass scale is much larger than the mass differences, and the allowed region discussed above has not yet split into distinct regions for the normal and inverted hierarchies. The last-published Cuoricino limit was $T^{0\nu}_{1/2}\geq3.1\times10^{24}\mbox{ yr}$ (90\% C.L.)~\cite{Arnaboldi}. However, the current Cuoricino limit (see Fig.~\ref{fig:limit}), a preliminary result including more statistics and using an updated Q-value measurement (see section~\ref{sec:detector}) as compared to the last-published result, revises this number to $T^{0\nu}_{1/2}\geq2.94\times10^{24}\mbox{ yr}$ (90\% C.L.)~\cite{Ferroni}; this corresponds to $\langle m_{\nu}\rangle\leq(210-700)\mbox{ meV}$, where the range arises from the different calculations of the nuclear matrix elements as tabulated in Ref.~\cite{Rodin}. There is also a claim to have actually seen $0\nu\beta\beta$ from a subset of the Heidelberg-Moscow collaboration (the Klapdor-Kleingrothaus claim), also in the degenerate-hierarchy region~\cite{Klapdor}. It is particularly important to test this claim using a different $\beta\beta$ isotope, with different systematics and nuclear matrix element calculations; when using a given matrix element calculation method, Cuoricino's results do not exclude the Klapdor-Kleingrothaus claim \cite{Arnaboldi}. Thus the goal of CUORE (the successor to Cuoricino), and indeed of all next-generation $0\nu\beta\beta$ experiments, is twofold: to test the Klapdor claim, and to extend sensitivity into the inverted-hierarchy region.

\section{Moving from Cuoricino to CUORE}

The CUORE experiment will be a search for evidence of neutrinoless double beta decay. It will use the bolometric technique established with its predecessor, Cuoricino, and should be able to achieve approximately a two-orders-of-magnitude improvement in sensitivity to the $0\nu\beta\beta$ half-life $T^{0\nu}_{1/2}$ over Cuoricino's current limit; CUORE's predicted limit after about five years of running is $T^{0\nu}_{1/2}\leq2.1\times10^{26}\mbox{ yr}$ (90\% C.L.), corresponding to $\langle m_{\nu}\rangle\leq(24-83)\mbox{ meV}$, assuming the $0.01\mbox{ counts}/[\mbox{keV}\cdot\mbox{kg}\cdot\mbox{y}]$ goal for the background rate is met (see discussion in section~\ref{sec:backgrounds} below).

\subsection{The Cuoricino and CUORE Detectors}
\label{sec:detector}

\begin{figure}[t]
 \centering
 \includegraphics[width=80mm]{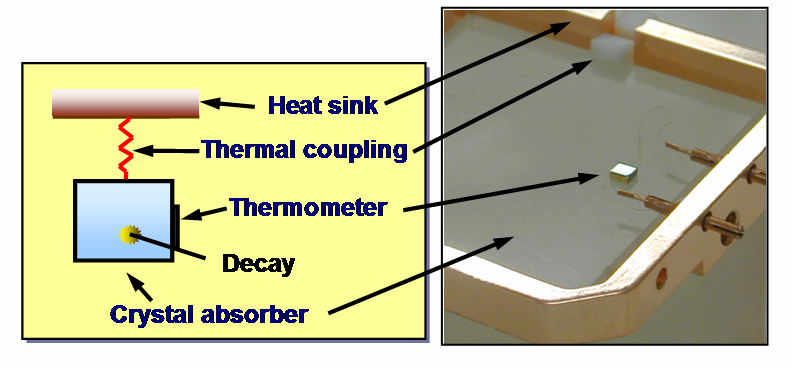}
 \caption{A side-by-side comparison of a schematic bolometer with a close-up photograph of the top face of a Cuoricino crystal mounted in its copper support structure. The chip glued to the crystal is the thermistor; the white block labeled in the figure as the `thermal coupling' is a PTFE standoff.}
 \label{fig:bolometer}
\end{figure}

The design of the CUORE detector is based on the bolometric principle (see Fig.~\ref{fig:bolometer}), as tested in the pilot experiment, Cuoricino. The detector is an array of (dielectric and diamagnetic) TeO$_{2}$ crystals, so that the detector itself contains the source ($^{130}\mbox{Te}$) that it is studying. Let $C$ be the heat capacity of one crystal, which is thermally coupled to a heat sink of temperature $T$ by a thermal conductance of $G$. Under the assumption that the crystal is a perfect calorimeter and $\Delta T(t)\ll T$ for all times $t$, whenever an event occurs in a crystal, the energy $\Delta E$ it deposits is converted to phonons and manifests as a small temperature rise
\begin{equation}
 \Delta T(t)=\frac{\Delta E}{C}e^{\left(-t/\tau\right)}
\end{equation}
\noindent
with recovery time constant
\begin{equation}
 \tau=\frac{C}{G}\mbox{ .}
\end{equation}
\noindent
According to the Debye law, at low temperature $C\propto$~$T^{3}$; therefore, the array is housed in a cryostat which keeps the crystals at 8-10~mK, reducing their heat capacity and thereby improving both the magnitude of the temperature response and the recovery time constant~\cite{Ardito}. Each crystal is equipped with a thermometer, or, to be more specific, a thermistor - a resistor whose resistance changes rapidly with temperature:
\begin{equation}
 R(T)=R_{0}e^{\sqrt{T_{0}/T}}
\end{equation}
\noindent
A bias current $I$ is applied across the thermistor, and the change in the resultant voltage $V$ across it as the thermistor's resistance changes with temperature is read out. The detector's exact response depends on the point on its $V-I$ load curve at which it is operating, known as the working point, which in turn depends on the base temperature of the bolometer. The thermistors therefore produce voltage data, which must be calibrated to signals of known energies to obtain a voltage-to-energy conversion.

\begin{figure}[t]
 \centering
 \includegraphics[width=15mm]{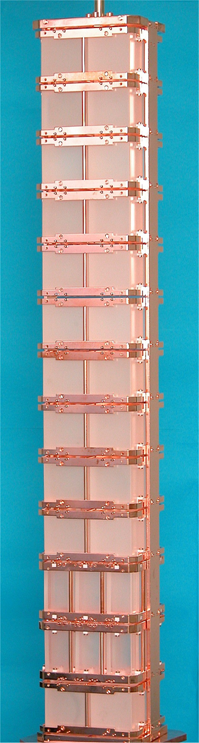}
 \hspace{5mm}
 \includegraphics[width=52mm]{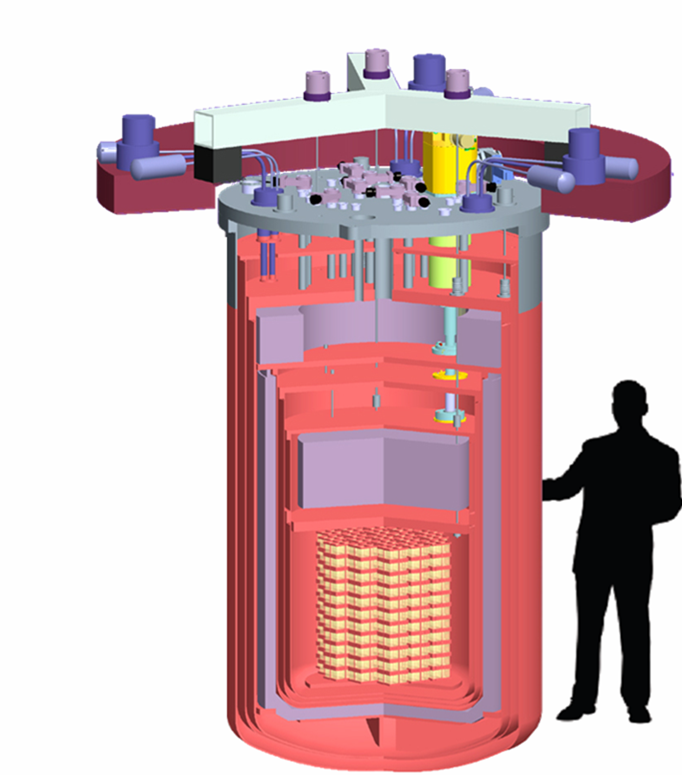}
 \caption{Left: A photo of the Cuoricino detector tower. Right: A 3D scale model of the CUORE detector (an array of 19 Cuoricino-like towers) and cryostat. The lavender structures are lead shielding.}
 \label{fig:detector}
\end{figure}

Both Cuoricino and CUORE use TeO$_{2}$ crystals to study the decay of $^{130}\mbox{Te}$. Tellurium is an advantage in this instance because of the relatively high natural abundance (33.8\%) of the $0\nu\beta\beta$ candidate isotope, which means that enrichment is not necessary to achieve a reasonably large active mass. Also, the Q-value of the decay falls between the peak and the Compton edge of the 2615~keV gamma line of $^{208}\mbox{Tl}$, the highest-energy gamma from the natural decay chains; this leaves a relatively clean window in which to look for the signal. Two recent measurements give the Q-value as $2527.01\pm0.32$~keV~\cite{Scielzo} and $2527.518\pm0.013$~keV~\cite{Redshaw}, a marked improvement in precision over the previously-accepted value of $2530.3\pm2.0$~keV~\cite{Audi}. Cuoricino was comprised of a single tower of crystals, with a total detector mass of 40.7~kg and a $^{130}\mbox{Te}$ mass of 11.34~kg. By contrast, CUORE will consist of an array of 988 $5\times5\times5\mbox{ cm}^{3}$ TeO$_{2}$ bolometer crystals, arranged in 19 towers of $2\times2\times13$ crystals apiece; the total detector mass will be 741~kg, which corresponds to a $^{130}\mbox{Te}$ mass of 203~kg.

The design and construction of the cryostat that will be used to maintain the detectors at the necessary cryogenic temperatures is a rather unique undertaking. It is based on the comparatively recently-developed technology of the cryogen-free dilution refrigerator, which utilizes pulse tube (PT) pre-cooling instead of a liquid helium bath; this should allow improved stability of the base temperature of the detectors as compared to the traditional $^{3}\mbox{He}/^{4}\mbox{He}$ refrigerator (used for Cuoricino). It will be the first cryostat of its kind big enough to house and cool the large detector mass represented by the CUORE array ($\sim$1~ton).

In addition to the increase in scale from Cuoricino to CUORE, in order for CUORE to reach its anticipated sensitivity, improvement is foreseen in two crucial aspects of detector performance: resolution and background. The average FWHM resolution at 2615 keV of the Cuoricino bolometers was approx.~8~keV; the goal for CUORE is 5~keV. Tests of the first batches of crystals produced for CUORE have shown that this goal has already been met, by means of improvements in the crystal quality, detector mounting structure, and reproducibility of the thermistor-crystal couplings. The average flat background in the region-of-interest seen in Cuoricino was $0.18\mbox{ counts}/[\mbox{keV}\cdot\mbox{kg}\cdot\mbox{yr}]$; the goal for CUORE is to reach $0.01\mbox{ counts}/[\mbox{keV}\cdot\mbox{kg}\cdot\mbox{yr}]$. See section~\ref{sec:backgrounds} below for further discussion of Cuoricino backgrounds and efforts to reduce them for CUORE. Projections of CUORE's sensitivity generally assume 5~keV resolution and $0.01\mbox{ counts}/[\mbox{keV}\cdot\mbox{kg}\cdot\mbox{yr}]$ background; an optimistic background rate of $0.001\mbox{ counts}/[\mbox{keV}\cdot\mbox{kg}\cdot\mbox{yr}]$ has also been considered.

\subsection{Backgrounds}
\label{sec:backgrounds}

$0\nu\beta\beta$ experiments must be low-background experiments, and CUORE is no exception. Backgrounds are the crucial limiting factor which controls the sensitivity which can be reached, and they must therefore be reduced and controlled as much as possible. The CUORE detector, like Cuoricino before it, will be located underground in the Laboratori Nazionali del Gran Sasso (LNGS) in Italy in order to reduce the rate of cosmic ray events; the cryostat will also contain shielding constructed from ancient low-radioactivity lead (see Fig.~\ref{fig:detector}) and be surrounded by additional lead to block environmental radioactivity from reaching the detector, and the detector structure itself will be composed of low-background materials and handled entirely in clean room conditions.

In Cuoricino, the main sources of the $0.18\pm0.01\mbox{ counts}/[\mbox{keV}\cdot\mbox{kg}\cdot\mbox{y}]$ background were the natural uranium and thorium decay chains from contamination of the detector materials. There were two main components to this background: surface contamination of the detector components, and bulk contamination of the cryostat materials. The surface contamination produced a flat $\alpha$ background in the region of interest; the main contributors were the surfaces of the copper support structures facing the bolometers (responsible for $50\pm20\%$ of the total background in the $0\nu\beta\beta$ region) and of the crystals themselves ($10\pm5\%$). The principal background contribution due to bulk contamination was the tail of the 2614.5~keV gamma produced by the decay of $^{232}\mbox{Th}$ in the cryostat materials ($30\pm10\%$). These values have been verified by extensive Monte Carlo studies~\cite{Arnaboldi}.

The goal that CUORE is striving to reach is $0.01\mbox{ counts}/[\mbox{keV}\cdot\mbox{kg}\cdot\mbox{y}]$. In order to reach this goal, more stringent material selection, production, cleaning, handling, and storage procedures have been established for all detector components to be used in the construction of CUORE. The cleaning of the copper support structures in particular has been the subject of an intense R\&D program which is currently in its final stages. Thus far, test runs have demonstrated background levels within a factor of 2-4 of the goal when extrapolated to CUORE.

To some extent, the array can self-veto against penetrating background particles like muons (which would cause simultaneous events in multiple adjacent crystals), and spurious events due to detector noise can be filtered out through pulse-shape analysis; however, in the end, the only real data, and therefore the only real handle on event identification, that bolomoters provide is energy information. This is quite suitable for a $0\nu\beta\beta$ experiment, since an energy signal is precisely what is sought, but it does mean that reliable, precise energy calibration is absolutely essential to the experiment's ability to provide meaningful data.

\section{The CUORE Detector Calibration System}

\begin{figure}[b]
 \centering
 \includegraphics[width=80mm]{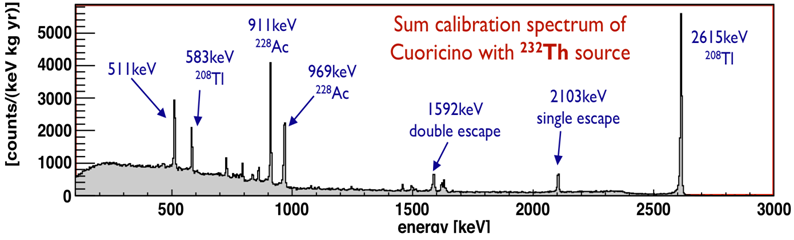}
 \caption{The summed spectrum of all Cuoricino bolometers, from a calibration run with $^{232}\mbox{Th}$. Each bolometer has been calibrated individually, then summed together after the calibration was applied. Note the strong 2615~keV gamma line from the decay of $^{208}\mbox{Tl}$, which provides a solid handle on the calibration near the $0\nu\beta\beta$~Q-value, 2530 keV.}
 \label{fig:cuoricinocalib}
\end{figure}

Since the thermistors read out the temperature changes in the bolometer crystals as voltage data, energy calibration must be performed in order to determine the relationship between the energy deposited in the crystal and the voltage signal subsequently obtained. This must be done for each bolometer individually, before the spectra can be summed together and analyzed for evidence of $0\nu\beta\beta$. In order to do this, a gamma source with a known spectrum is used to illuminate the crystals. Although the most critical energy region that must be calibrated is the region of interest around the $0\nu\beta\beta$~Q-value, the whole spectrum should be calibrated as well as possible for reliable identification of backgrounds. For this purpose, Cuoricino used $^{232}\mbox{Th}$ as its calibration source, since its decay chain produces a number of gamma lines up to 2615~keV that are strong enough to be used for calibration; the 2615~keV peak is particularly strong, and can be used to ensure solid calibration in the region of interest (see Fig.~\ref{fig:cuoricinocalib}). The calibration uncertainty in the region of interest is a systematic error in the determination of $T^{0\nu}_{1/2}$. In Cuoricino, this uncertainty was negligible with respect to the $\pm2$~keV uncertainty on the Q-value; however, with the improved precision in recent Q-value measurements, the performance of the CUORE calibration system is of greater importance.

\begin{figure*}[ht]
 \centering
 \includegraphics[width=135mm]{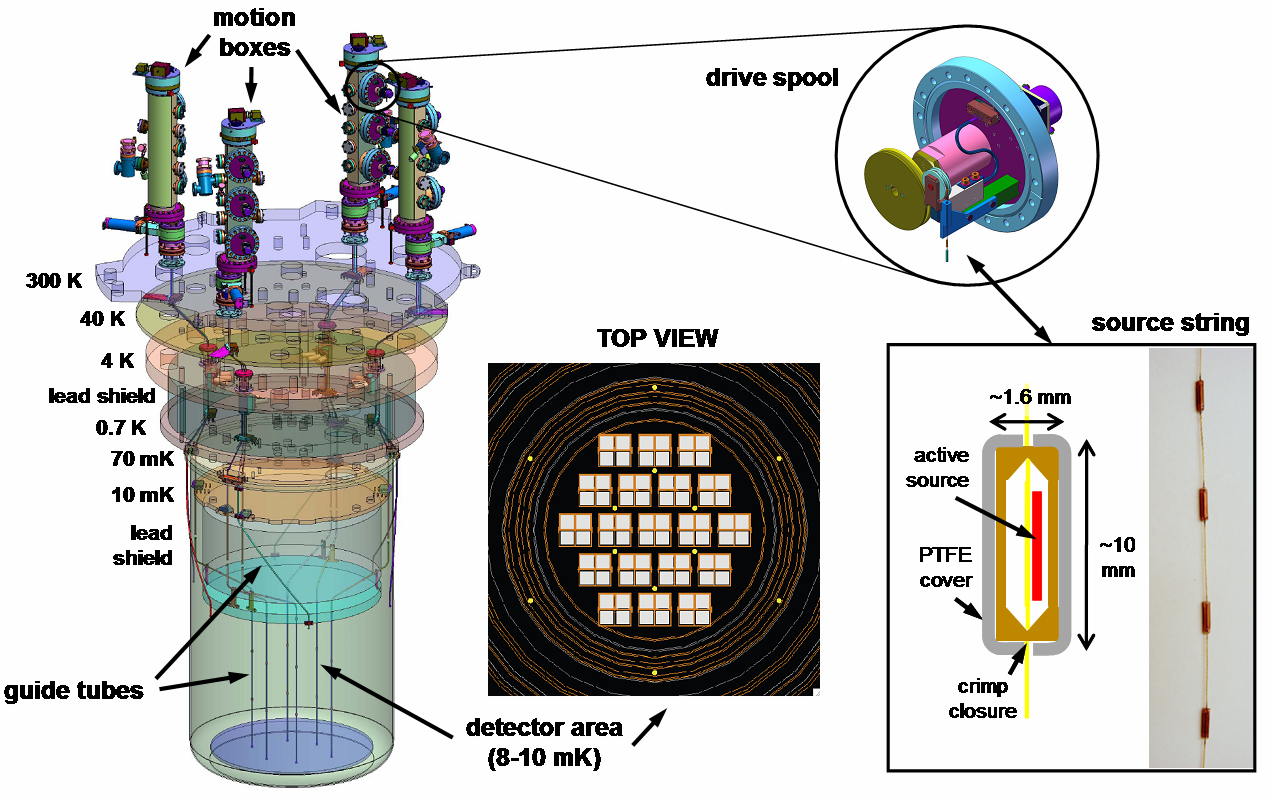}
 \caption{Some images illustrating the calibration system concept. Left: 3D model of the calibration system as it interfaces with the cryostat flanges and vessels. Center: source positions relative to detector towers and cryostat shields. Top right: 3D model of the drive spool (as seen from inside the motion box). Bottom right: schematic of a single source capsule; photograph of a section of a prototype source string.}
 \label{fig:overview}
\end{figure*}

The thermistors which are used to read out the crystals are very sensitive and their response will change with small variations in the working point of the detector. Between calibrations, the response of the bolometers is stabilized by means of periodic heater pulses of known energy, and the base temperature of the detector is stabilized with a DC feedback loop; however, it is still necessary to perform a calibration every month or two with minimal disruption of the detector and cryostat. `Minimal disruption' means that the calibration system must not compromise the low-background environment of the detector, nor place excessive thermal load on the cryostat such that the detector warms up, changing the working points of the bolometers.

One of the most dramatic changes that has to be made in scaling up from the single tower of Cuoricino to the 19 towers of CUORE is that the calibration system must grow in complexity. In Cuoricino, which comprised a single tower of crystals, two radioactive source wires were inserted inside the external shielding on either side of the cryostat. In CUORE, the outer towers will shield the inner towers, so some calibration sources must be routed all the way into the detector area and between the towers (see bottom-center image in Fig.~\ref{fig:overview}) in order to achieve even illumination, which is important in order to be able to successfully calibrate the entire detector in a reasonable amount of time without causing an excessively high event rate in some crystals - since bolometers are inherently slow (each pulse lasts several seconds), a high rate causes pileup and raises the baseline temperature of the detectors, leading to increased dead time and degradation of the energy resolution.

The CUORE detector calibration system addresses this requirement in the following way: the system consists of 12 flexible source carriers, routed through the levels of the cryostat by means of guide tubes, and stored and deployed by four motion boxes containing three spools each which sit on top of the 300~K flange of the cryostat. This approach allows the sources to be stored entirely outside the cryostat during normal data-taking, and to traverse the complicated routes through the interior of the cryostat (see Fig.~\ref{fig:overview}) necessary to reach the detector area. Motion in cryogenic and vacuum conditions is challenging, because of the mechanical and thermal effects of friction and vibration; additional complication arises from the fact that the calibration sources must travel through regions of differing temperatures, from 300~K to 8-10~mK, without failing under thermal cycling or thermally overloading the cryostat. The system has been carefully designed to meet this challenge; a more detailed discussion than that which follows can be found in~\cite{Sangiorgio}.

The source carrier is conceived as a collection of small, individual active sources, chained together to form a single flexible unit that is capable of sliding down through the guide tubes into calibration position under its own weight when fed off a spool (see Fig.~\ref{fig:overview}). The source carrier will be built by attaching individual source capsules to a continuous string. A source capsule is comprised of a copper tube crimped to the string, and covered with PTFE heat-shrink tubing to reduce the friction against the guide tube during source motion. Each capsule will house a length of thoriated tungsten wire. The source isotope will thus be $^{232}\mbox{Th}$, as it was in Cuoricino; however, this source carrier design is adaptable enough to allow the possibility of using different isotopes in the case that it is deemed useful to do so.

The guide tubes route the source carriers through the cryostat, and also provide a thermal connection to various stages of the cryostat. Heat will be dissipated into the cryostat by the friction of the source carriers moving through the bends. Additionally, since the guide tubes essentially form a penetration through all the thermal stages of the cryostat, the thermal gradient that will develop along them represents another thermal load; the materials for the guide tubes must therefore be chosen on the basis of both thermal conductivity, to minimize thermal load, and radiopurity, to minimize background events induced in the detector. As a conservative approach, the system has been designed to operate within the static heat budget of the cryostat wherever possible, even during source motion, since we do not know the dynamic recovery time constant of which the cryostat will be capable.

Calculations have shown that, in order for the thermal radiation from the source carriers to be at an acceptable level when they are in the detector area, the source carriers must be thermalized to 4~K during the insertion process. A thermalization clamp has been designed in order to ensure sufficient thermal contact between the sources and the guide tubes; four of these clamps will be mounted below the 4~K flange of the cryostat, one for each group of three guide tubes.

Extensive room-temperature motion, friction, and control tests have been conducted of a prototype guide tube system. Motion has been shown to be reliable, and instrumentation (including a motor encoder, a proximity sensor, and a load cell operated as a tension meter) allows monitoring of the sources' travel through the tubes and automatic failsafes against the source string escaping the tubes or damage to the motor. Vacuum and cold tests of the system will be conducted as part of the commissioning of the CUORE cryostat.

\section{Conclusions and Status of the Experiment}

CUORE is in the construction phase. The facilities at LNGS are under construction and more than 200 crystals have been received and stored at LNGS. In early 2010, assembly and installation of the first CUORE tower in the Cuoricino cryostat will begin; this tower will take data independently as CUORE-0, providing an `engineering run' to verify CUORE assembly procedures as well as allowing statistics compatible with those of Cuoricino to continue to accrue until the full CUORE array is operational. Also in early 2010, assembly of the CUORE cryostat will begin, allowing a battery of hardware tests at both room temperature and at cold (including cold tests of the calibration system) to be performed. CUORE detector construction will continue throughout 2010 and 2011; data taking with the full CUORE array will begin in 2012.

CUORE is a next-generation $0\nu\beta\beta$ experiment which will utilize the bolometric detection technique proven in its predecessor, Cuoricino. It is now in the construction phase, and will be one of the first $0\nu\beta\beta$ experiments to probe the inverse hierarchy mass region.


\bigskip 

\begin{thebibliography}{99} 

\bibitem{Barabash}A. S. Barabash, Czech J. Phys. \textbf{56}, 437 (2006).

\bibitem{Elliot}Steven R. Elliot and Petr Vogel, Ann. Rev. Nucl. Part. Sci. \textbf{52}, 115 (2002); C. Aalseth \textit{et al.}, arXiv:hep-ph/0412300v1 (2004).

\bibitem{McKeown}R.D. McKeown and P. Vogel, Phys. Rept. \textbf{394}, 315 (2004).

\bibitem{Schechter}J. Schechter and J. Valle, Phys. Rev. D \textbf{25}, 2951 (1982).

\bibitem{Piquemal}Fabrice Piquemal, J. Phys. Conf. Ser. \textbf{120}, 052004 (2008); A. Strumia and F. Vissani, preprint arXiv:hep-ph/0606054v2 (2007).

\bibitem{Arnaboldi}C. Arnaboldi \textit{et al.} [CUORICINO collaboration], Phys. Rev. C \textbf{78}, 035502 (2008).

\bibitem{Ferroni}F. Ferroni, \textit{To be published in the proceedings of Rencontres de Moriond EW 2009, La Thuile, Italy, March 7-14}.

\bibitem{Rodin}V. A. Rodin, A. Faessler, F. Simkovic, and P. Vogel, Nucl. Phys. \textbf{A766}, 107 (2006); [Erratum Nucl. Phys. \textbf{A793}, 213 (2007)].

\bibitem{Klapdor}H.V. Klapdor-Kleingrothaus et al., Mod. Phys. Lett. A \textbf{16}, 2409 (2001); H.V. Klapdor-Kleingrothaus et al., Nucl. Inst. and Meth. A \textbf{522}, 371 (2004).

\bibitem{Ardito}R. Ardito \textit{et al.} [CUORE collaboration], arXiv:hep-ex/0501010v1 (2005).

\bibitem{Scielzo}N. D. Scielzo \textit{et al.}, Phys. Rev. C \textbf{80}, 025501 (2009).

\bibitem{Redshaw}Matthew Redshaw, Brianna J. Mount, Edmund G. Myers, and Frank T. Avignone III, Phys. Rev. Lett. \textbf{102}, 212502 (2009).

\bibitem{Audi}G. Audi, A.H. Wapstra, and C. Thibault, Nucl. Phys. \textbf{A729}, 337 (2003).

\bibitem{Sangiorgio}S. Sangiorgio \textit{et al.}, \textit{To be published in the proceedings of the 13th International Workshop on Low Temperature Detectors (LTD), Stanford, CA, July 20-24}, preprint arXiv:nucl-ex/0908.0167v1 (2009).

\end{thebibliography}

\end{document}